\documentclass[10pt,twocolumn,letterpaper]{article}

\usepackage{cvpr}
\usepackage{times}
\usepackage{epsfig}
\usepackage{graphicx}
\usepackage{amsmath}
\usepackage{amssymb}
\usepackage{subcaption}
\usepackage{amsmath, siunitx}
\usepackage{booktabs}

\usepackage{ifthen}
\newboolean{ack}
\setboolean{ack}{true}

\newboolean{combined}
\setboolean{combined}{true}
\newcommand{\beginsupplement}{%
        \setcounter{table}{0}
        \renewcommand{\thetable}{S\arabic{table}}%
        \setcounter{figure}{0}
        \renewcommand{\thefigure}{S\arabic{figure}}%
        \setcounter{section}{0}
        \renewcommand{\thesection}{S\arabic{section}}%
}

\usepackage[pagebackref=true,breaklinks=true,letterpaper=true,colorlinks,bookmarks=false]{hyperref}

\cvprfinalcopy 

\def\cvprPaperID{23} 

\ifcvprfinal\fi

\begin{document}

\title{Calibrated Vehicle Paint Signatures for Simulating Hyperspectral Imagery}

\author{Zachary Mulhollan\\
Rochester Institute of Technology\\
{\tt\small zjm1400@rit.edu}
\and
Aneesh Rangnekar\\
{\tt\small aneesh.rangnekar@mail.rit.edu}
\and
Timothy Bauch\\
{\tt\small tdbpci@cis.rit.edu}
\and
Matthew J. Hoffman\\
{\tt\small mjhsma@rit.edu}
\and
Anthony Vodacek\\
{\tt\small axvpci@rit.edu}
}

\maketitle

\begin{abstract}
   We investigate a procedure for rapidly adding calibrated vehicle visible-near infrared (VNIR) paint signatures to an existing hyperspectral simulator - The Digital Imaging and Remote Sensing Image Generation (DIRSIG) model - to create more diversity in simulated urban scenes. The DIRSIG model can produce synthetic hyperspectral imagery with user-specified geometry, atmospheric conditions, and ground target spectra. To render an object pixel's spectral signature, DIRSIG uses a large database of reflectance curves for the corresponding object material and a bidirectional reflectance model to introduce s due to orientation and surface structure. However, this database contains only a few spectral curves for vehicle paints and generates new paint signatures by combining these curves internally. In this paper we demonstrate a method to rapidly generate multiple paint spectra, flying a drone carrying a pushbroom hyperspectral camera to image a university parking lot. We then process the images to convert them from the digital count space to spectral reflectance without the need of calibration panels in the scene, and port the paint signatures into DIRSIG for successful integration into the newly rendered sets of synthetic VNIR hyperspectral scenes.
\end{abstract}

\section{Introduction}

\begin{figure}[t]
    \centering
    \includegraphics[width = \linewidth]{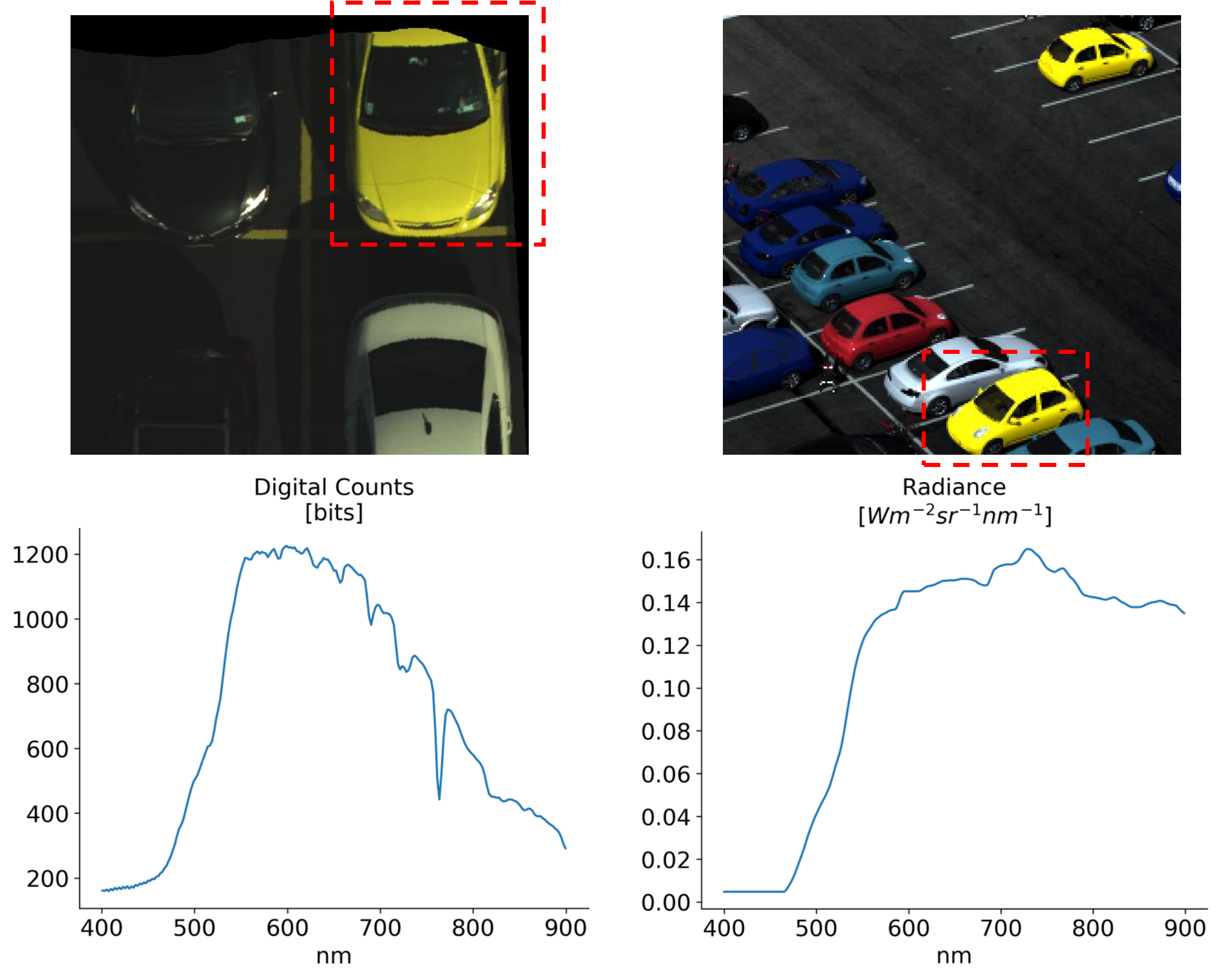}
    \caption{Examples of real and simulated hyperspectral data from a parking lot (displayed as RGB). Left, an image chip from the flight line (top) and the VNIR spectrum of digital counts obtained from a region of interest sampled from the yellow car (bottom). Right, a synthetic parking lot scene with cars exhibiting several paint signatures (top) including the radiance spectrum sampled from a yellow car (bottom).}
    \label{fig:frontpage}
\end{figure}

The availability of large scale datasets has been an important factor in the ongoing success of deep neural networks - for example, the Mask R-CNN framework for image instance segmentation \cite{imagenet_cvpr09,lin2014microsoft,he2017mask}. Computer vision algorithms often use synthetic data for training and testing deep neural networks and then adapt them to the real data distribution for application deployments \cite{wymann2015torcs,dosovitskiy2017carla,wrenninge2018synscapes,savva2019habitat}. In comparison, research in hyperspectral imagery for these kind of tasks is limited due to the cost of hardware and acquisition of a wide variety of object signatures for input to synthetic data simulators. Recently, AeroRIT established a baseline on the task of semantic segmentation with hyperspectral imagery \cite{rangnekar2019aerorit} by annotating all pixels in a flight line over a university campus. The authors show trained networks being able to correctly identify vegetation, buildings and road pixels to a large degree while struggling with cars due to the relatively low resolution and the atmospheric noise in the scene. We hypothesize that synthetic imagery can help improve the performance of such networks by providing better initialization and a larger set of training samples. After exploring the available simulators for hyperspectral scene rendering, we find that current simulators are geared more towards vegetation and cloud modeling, with few samples for vehicle paints. To increase the number of available paint samples, we fly a drone over a parking lot filled with cars and devise a methodology to extract the digital count signatures from the vehicles for modeling a new set of paint signatures (Fig. \ref{fig:frontpage}). 

\begin{figure*}[]
    \begin{subfigure}{\textwidth}
        \centering
        \includegraphics[width=\linewidth]{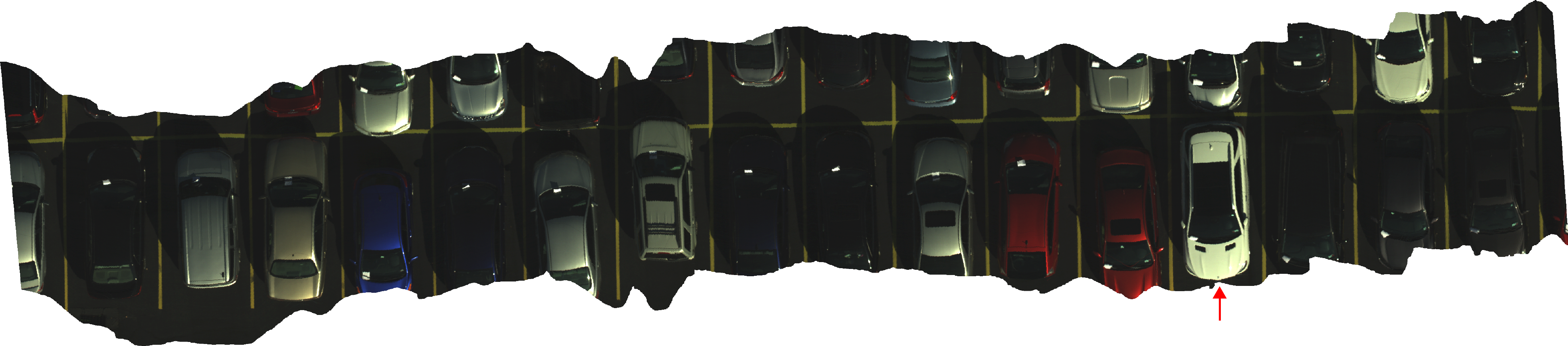}
        \caption{}
        \label{fig:exampleflightline}
    \end{subfigure}
    
    \begin{subfigure}{0.32\textwidth}
        \centering
        \includegraphics[width=\linewidth]{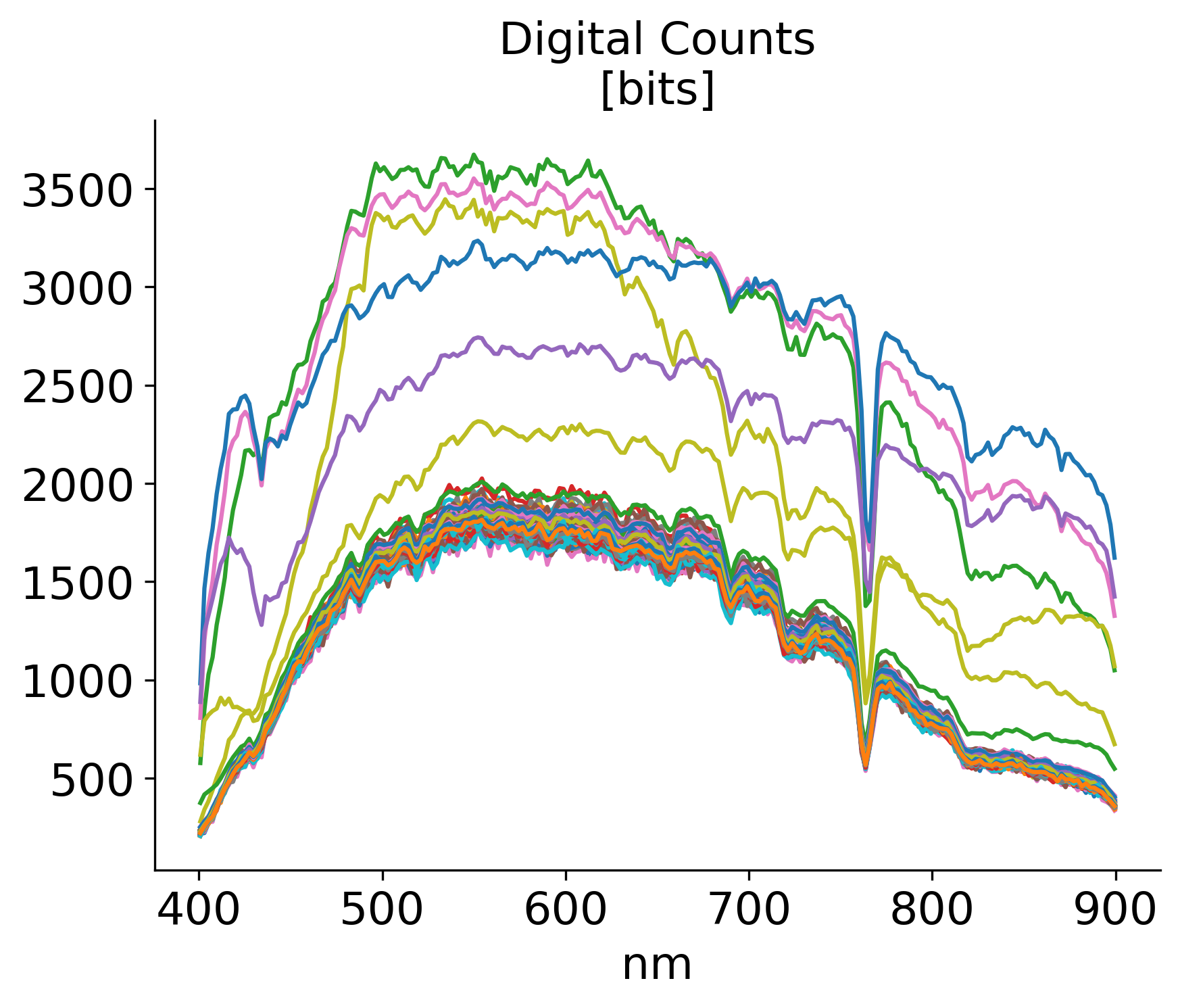}
        \caption{}
        \label{fig:car_dc}
    \end{subfigure}
    \hfill
    \begin{subfigure}{0.32\textwidth}
        \centering
        \includegraphics[width=\linewidth]{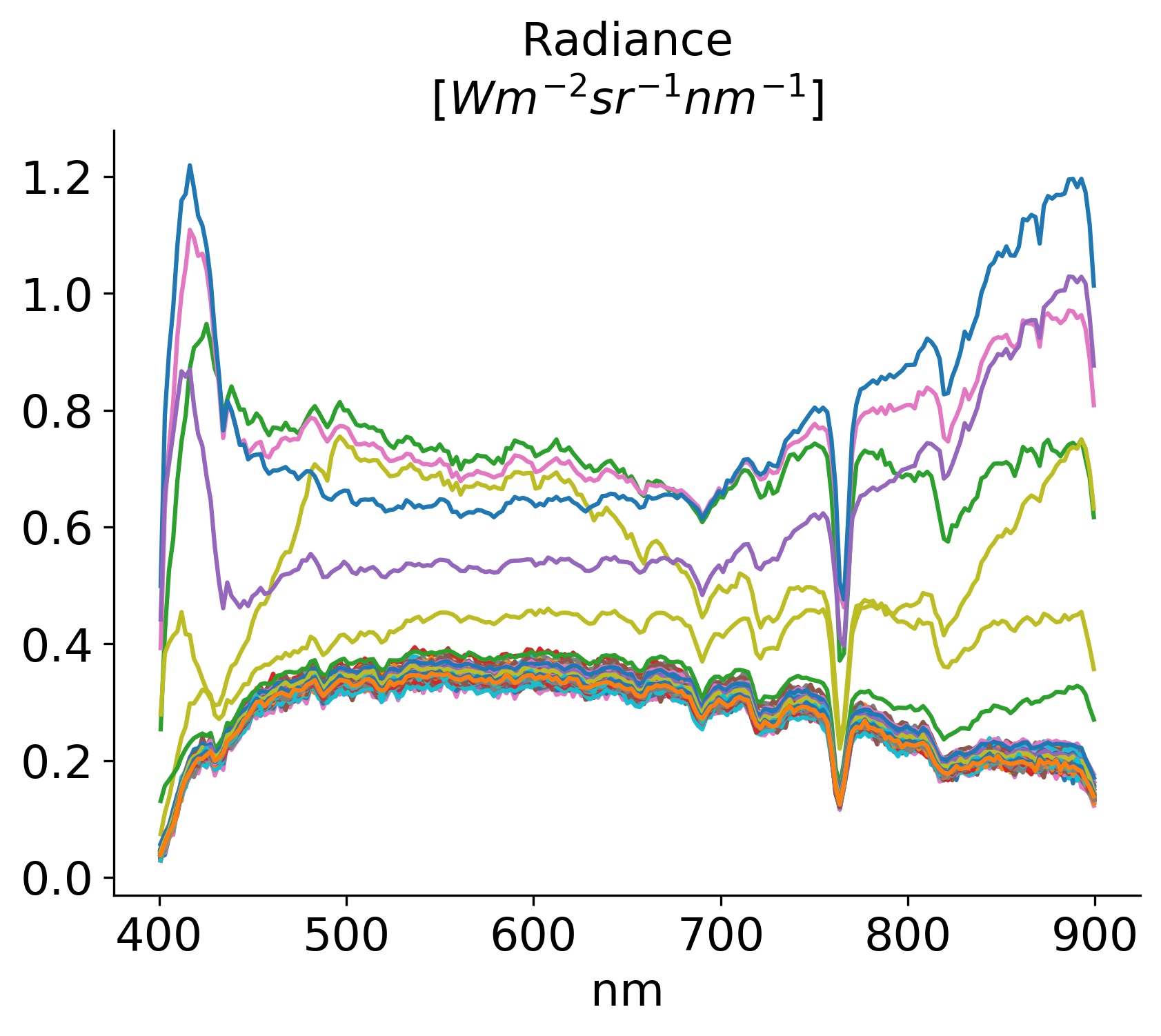}
        \caption{}
        \label{fig:car_rad}
    \end{subfigure}
    \hfill
    \begin{subfigure}{0.32\textwidth}
        \centering
        \includegraphics[width=\linewidth]{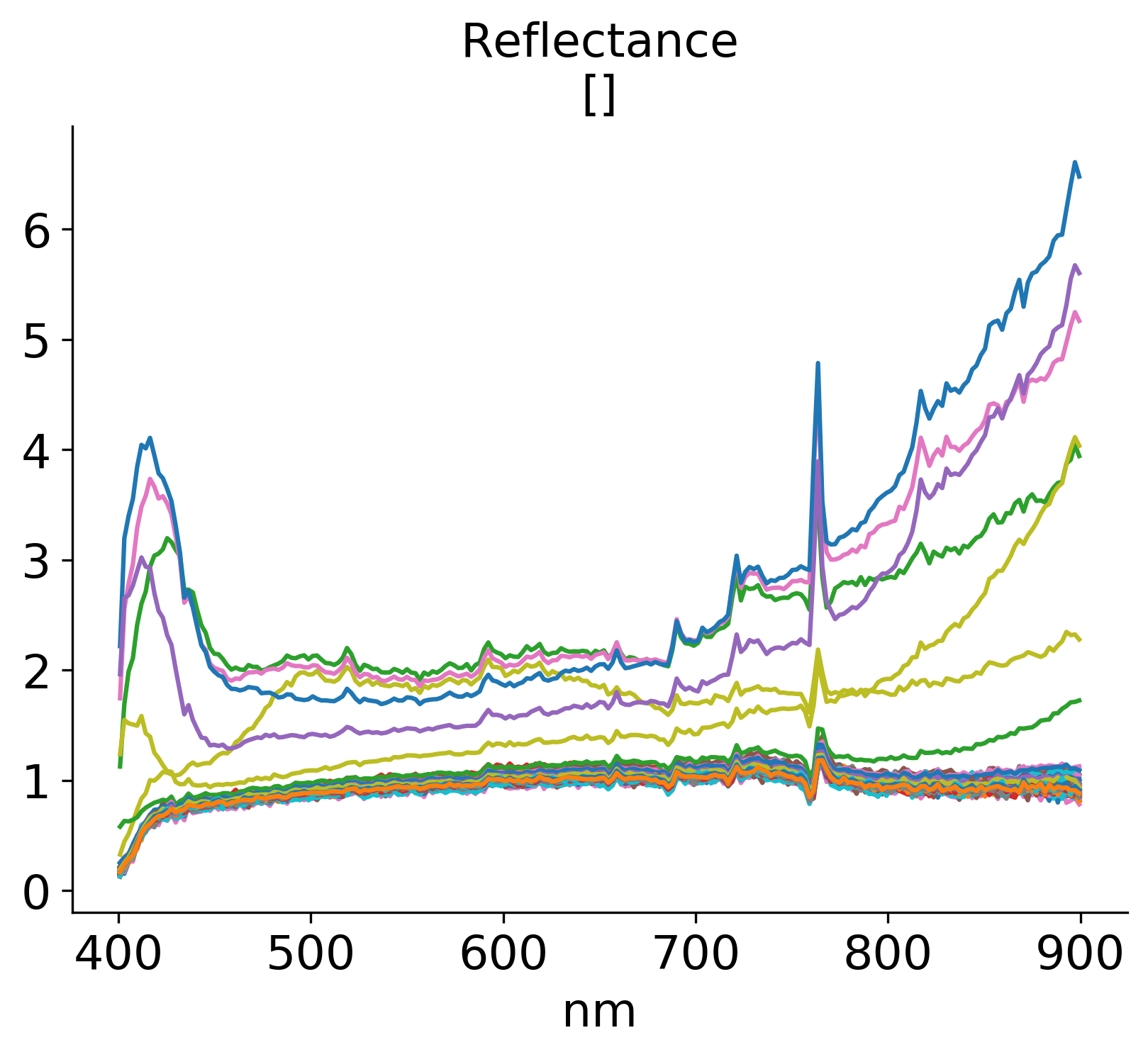}
        \caption{}
        \label{fig:car_ref}
    \end{subfigure}
    
    \caption{\ref{fig:exampleflightline} shows a few vehicles captured from the hyperspectral sensor during our data collection. We obtain a rich dataset consisting of diverse paints and their sub-variants under different illumination conditions. The wavy image edges are a feature of the orthorectification process. \ref{fig:car_dc}, \ref{fig:car_rad}, \ref{fig:car_ref} show the high level conversion from digital counts to radiance to reflectance for a few pixels from the vehicle of interest pointed by the red arrow (off-white SUV). The neatly clustered set of curves indicate the vehicle paint spectra, and the rest of the curves indicate glint. Reflectance is unitless.}
    \label{fig:Bigfig}

\end{figure*}

Four well-known simulators that can render hyperspectral scenes are: MCScene \cite{Richtsmeier2001A3R}, CameoSim \cite{moorhead2001cameo}, DIRSIG \cite{dirsig1, goodenough2017dirsig5}, and CHIMES \cite{zahidi2020end}. MCScene (Monte Carlo Scene) generates the imagery by modeling a 3D cuboid world with all the specified atmospheric parameters and then using Direct Simulation Monte Carlo for all spectral signatures. CameoSim (CAMoflauge Electro-Optic Simulator) represents objects in the scene as polygons and then applies radiosity and reverse ray tracing to model the signatures. The DIRSIG (Digital Imaging and Remote Sensing Image Generation) model uses physics-based radiation propagation modules along with Metropolis Light Transport for scene rendering. CHIMES (Cranfield Hyperspectral Image Modeling and Evaluation System) is relatively new and uses an enhanced adjacency model and automatic atmospheric parameter search to render realistic scenes with minimal user effort. We use DIRSIG as the simulator following the extensive list of successful works \cite{kolb2016digital,han2017efficient,rahman2018siamese,uzkent2016integrating,uzkent2018tracking,kemker2018algorithms}.

Kolb \etal used DIRSIG to create virtual night-time scenes for conducting system tests over a wide range of atmospheric visibility and environmental conditions \cite{kolb2016digital,meyers2002incorporation}. Han \etal modeled urban and desert scenes to create a large dataset of synthetic remote sensing images of cars and helicopters under a wide combination of atmospheric and environmental factors for data augmentation to train convolutional neural networks \cite{han2017efficient}. Rahman \etal rendered a dataset similar to Han \etal and used a VGG16-based siamese network for change detection under various illuminations \cite{rahman2018siamese,simonyan2014very, bromley1994signature}. Uzkent \etal modeled vehicle movements observed by imaging platforms at different altitudes for the purpose of detection and tracking using the Megascene environment in DIRSIG \cite{ientilucci2003advances,uzkent2016integrating,uzkent2018tracking}. Kemker and Kanan used the Trona scene from DIRSIG to boost the average accuracy of their ResNet-50 based semantic segmentation architectures \cite{kemker2018algorithms, he2016deep}.

The current version of DIRSIG supports fewer than 10 paint signatures for vehicles. These spectra can then be added to synthesize new paint signatures and create the entire color gamut in the 3 band RGB domain. However, simulating realistic hyperspectral imagery requires a large database of spectral color signatures for each desired material, such that the natural  of a material's spectra can be measured and then rendered in simulated environments. To our knowledge, such a dataset for vehicle paint spectra is not publicly available. Hence, we collect the required paint data and augment the existing set of signatures for a wider range of samples. We discuss our approach towards this objective in the paper.

Our contributions are summarized as follows: 
\begin{itemize}
    \item We introduce a relatively simple approach to process data in digital counts into reflectance, without the explicit need of calibration panels.
    \item We gather a rich dataset of car paints that is much larger than the current core car paint signatures in DIRSIG, with metadata containing the car make and model for future use. 
\end{itemize}

\section{Data Collection}
\label{sec:Data_Collection}

We collect a comprehensive, high-resolution vehicle dataset on a university parking lot to obtain hyperspectral signatures of different vehicle paints (Fig. \ref{fig:exampleflightline}). We start the collection at noon, when the parking lot is near full capacity with 450+ vehicles. We obtain ground truth spectra of 10 vehicles in the scene by measuring their hood with a hand-held SVC spectrometer for verifying the spectral curves obtained by our method. The main imaging platform for this collect is inspired by the setup of Kaputa \etal - a drone fitted with a push broom hyperspectral sensor \cite{kaputa2019mx}. We summarize all the instruments and specifications of the sensor and flight in Table \ref{tab:allspecs}.

\begin{table}[h]
\centering
\def\arraystretch{1.2}
\begin{tabular}{@{}ll@{}}\toprule
\textbf{Imaging System} \\
\midrule 
Manufacturer & Headwall Photonics\\
Model & Nano-Hyperspec \\
Spectral Range [nm] & 400 - 1000\\
Spectral Bands & 272 \\
Bit Depth (bits) & 12 \\
\midrule
\textbf{Other sensors} \\
\midrule 
Device & SVC Spectrometer\\
Spectral Range [nm] & 335 - 2510\\
Device & ASD Spectrometer\\
Spectral Range [nm] & 350 - 2500\\
\midrule
\textbf{Flight} \\ \midrule
Altitude [ft] & 50 \\
GSD [cm] & $\approx$ 0.8\\
Exposure time [ms] & 5 \\
Frame period [ms] & 7 \\
\bottomrule \\
\end{tabular}
\caption{Specifications for data collection.}
\label{tab:allspecs}
\end{table}

The \textbf{Nano-Hyperspec} is a pushbroom scanning sensor that collects a simultaneous line of 640 cross-track spatial pixels and 272 spectral band pixels. Along-track pixels are collected over time as the drone flies over the parked vehicles in the scene. We collect an average of 50,000 hyperspectral pixels per vehicle, providing a dense number of data points that captures the natural intra-vehicle material reflectance . We carefully sample regions from these set of points as described in Sec. \ref{sec:roiselection}.

The sources of light energy for this data collection are sunlight and skylight, which in the remote sensing field are termed downwelling irradiance.  It is important to accurately measure the downwelling irradiance because the source of light in a scene impacts the observed color or spectra of an object's surface. Atmospheric effects such as clouds, shadows, and haze, as well as the sun's position in the sky, need to be measured to calibrate raw hyperspectral data. To keep track of changes in the atmosphere and sun during the vehicle collect, we use an \textbf{ASD Spectrometer} pointed skyward at a stationary position on a grounded tripod. We attach an optically diffuse cosine corrector to the ASD spectrometer so that it collects light with a 180 degree field of view, which provides our dataset a measure of the full-sky hemisphere downwelling irradiance every 2 seconds.

Following in the order of the light path, the downwelled light then interacts with materials where the light either reflects, absorbs, or transmits. Our main interest is in measuring the reflective properties of vehicle materials to then generate hyperspectral paint signatures. We utilize a factory-calibrated handheld \textbf{SVC spectrometer} to measure spectral radiance in the VIS-SWIR range with 987 spectral bands. To convert radiance into reflectance, the SVC spectrometer uses a Spectralon target that has a flat reflectance curve ( $>$ 99\% reflective in VIS-SWIR range) and is highly Lambertian. We obtain the spectral reflectance of vehicle paints by dividing the radiance of the vehicle paint by the Spectralon target radiance and treat it as the ground truth reflectance data. Reflectances measured by the SVC spectrometer are used to verify our calculated reflectances from the Nano-Hyperspec as shown in Sec. \ref{sec:comparisonwithsvc}.



To summarize, we have the following sets of spectra from the data collection that we use to move from Fig. \ref{fig:car_dc} to Fig \ref{fig:car_ref}:
\begin{itemize}
    \item \textbf{Nano-Hyperspec:} Flight line data in digital counts domain.
    \item \textbf{ASD:} Downwelling irradiance throughout the data collection at 0.5 \si{\hertz}.
    \item \textbf{SVC:} Spectral radiance and reflectance for a selected set of cars we use for verification of our pipeline.
\end{itemize}

\section{Radiometric Calculations}
\label{sec:radcalc}

Collecting airborne hyperspectral imagery of vehicles and calibrating the data to surface material reflectance is a difficult task due to complex object geometry and radiation propagation. For our university parking lot dataset - a scenario that is starkly different than a controlled laboratory setup - it is necessary for us to make certain assumptions about the vehicle reflectance properties and the illumination sources. Even though vehicle paint is highly specular, we assume vehicle paint is Lambertian so that we can approximate spectral radiance from radiant exitance leaving the vehicle's surface without performing an exhaustive in-situ measure of the surface BRDF (bidirectional reflectance distribution function), as shown in Eqn. \ref{eq:Radiance_Exitance}. 

\begin{equation}
    L_\lambda = \frac{M_\lambda}{\pi} = \frac{E_\lambda \rho_\lambda \cos{\theta}}{\pi},
    \label{eq:Radiance_Exitance}
\end{equation}

where $L$, $M$, $E$, $\rho$ indicate radiance, radiant exitance, irradiance and reflectance at wavelength $\lambda$, and $\theta$ is the angle of incidence. We summarize the terms across all equations in Table \ref{tab:allunits}. When calculating vehicle reflectance, we select regions of interest on the car's body that are approximately planar and nadir to the drone imaging platform so that measured variance in exitance due to surface angle $\theta$ is reduced. 

For our dataset, we assume the primary source of illumination in the scene is from sky downwelling irradiance, and that all other sources of illumination are negligible. In local cases where an image pixel contains shadowed or glinted light from the sun, this assumption is no longer true and therefore calculated surface reflectance from these pixels will not return an accurate spectral curve, as demonstrated in Fig. \ref{fig:car_ref}. Additionally, since some vehicle paints are highly reflective and the vehicles are placed close to each other in our parking lot dataset, the adjacency effect (secondary reflections from nearby objects) can act as another significant illumination source. We have identified small regions of interest (ROI) in our dataset where the adjacency effect prominently alters our spectral data, which we discuss in detail in Sec. \ref{sec:roiselection}. 

Collecting surface material spectra remotely with a hyperspectral imaging system requires a method to calibrate the sensors electronic digital counts to physically meaningful units, such as radiance or reflectance. For airborne remote sensing at high altitude, accurate radiometric calibration of data to reflectance requires an accounting of all light-matter interactions in the atmosphere between a materials surface and the airborne sensor. Since the university parking lot data was collected with a drone flying at very low altitude (50 feet) on a clear and sunny day, we assume that the radiometric atmospheric effects between the sensor and ground are negligible.

The physical process in which a photosensitive sensor generates electrons when impinged by photons can be described as


\begin{equation}
    S_e = \phi \cdot t \cdot \frac{\lambda}{hc} \cdot \eta,
    \label{eq:sensor_electrons}
\end{equation}

where $\phi$ is radiant flux, $t$ is time, and $\eta$ is the ratio of converted electrons to incident photons per wavelength, also known as quantum efficiency (QE). When a sensor is photon-noise limited, that is, the photon Poisson noise is greater than the detectors inherent electronic noise, the quantum efficiency is the dominant factor in determining a sensor's maximum achievable signal to noise ratio (SNR). A sensor's QE can be measured in a controlled laboratory setting when using a monochromatic light source of known intensity, and recording the number of electrons generated by the sensor.

Extending beyond the sensor, the conversion of photons to electrons of an imaging system can be defined holistically with external quantum efficiency (EQE). EQE is the ratio of photons incident on the first optical element to electrons generated by the sensor. This can be a convenient method to define performance of an imaging system with complex static optical components such as lenses, apertures, mirrors, and diffraction gratings because EQE accounts for any potential loss in signal through the entire optical system.

A synonymous measure to QE and EQE is spectral responsivity, which is the ratio of current generated to incident radiometric power. For a camera, output pixel values are expressed in digital counts after the analog current in the sensor is passed through an analog-to-digital converter. Assuming the analog-to-digital converter and the sensor are linear, system spectral responsivity can be defined in terms of digital counts as 


\begin{equation}
    R_\lambda  = \frac{DC_i - DC_{dark}}{\phi}.
    \label{eq:responsivity}
\end{equation}

A benefit of measuring spectral responsivity in terms of camera digital counts is that, like EQE, it takes into account the affect of optical components as well as the photosensitive sensor on incident power. This contextually aligns with a system calibration process that is needed for collecting field data where the raw output is in digital counts. 

Normalizing the spectral responsivity curve by its maximum will provide a relative spectral responsivity for an imaging system. Dividing raw digital count data (that is corrected for dark current) by an accurately measured relative spectral responsivity curve will ensure that the incident power required to raise a pixel's value by one digital count will be constant for all spectral bands in the imaging system. For applications which do not require image pixel values to be in absolute physical units, further radiometric calibration may not be needed.

To convert a camera's digital counts to absolute physical units such as radiance, the straightforward method is to measure the camera's digital count response in a controlled laboratory setting, where the light source power spectrum and spectral bandwidth are known. In the specific case of calibrating the Nano-Hyperspec camera to spectral radiance, we used a monochrometer with a tunable diffraction grating. We placed the front of the Nano-Spec as close to the exit aperture of the monochrometer as possible and blocked out stray light with optical absorbing material. We collected Nano-Hyperspec image frames with monochromatic light every 10nm in the VIS-NIR range (61 steps from 400-1000nm). For each wavelength tested, we also measured the monochrometer's lamp radiant flux $\phi_{ref}$ with an optical power meter, which is used in Eqn. \ref{eq:MulhollanMethod}.

Since the Nano-Hyperspec has a diffraction grating to separate incoming light into 272 spectral bands across the image array columns, a monochromatic light source will only illuminate a few columns with the rest of the image array receiving almost no light. With a theoretical noiseless system, we could determine the spectral responsivity of the camera by recording the maximum digital count value in the illuminated band for each monochromatic wavelength tested, then use Eqn. \ref{eq:responsivity} to calculate responsivity. However, there is noise introduced from digitized sampling, non-uniform illumination, and spectral band widening due to diffracted light that needs to be addressed.

Light leaves the monochrometer aperture with non-uniform intensity, and across the illuminated band on the image plane the intensity profile is approximately Gaussian. For each monochromatic wavelength tested, we fit a 1-D Gaussian function to our data and record the amplitude in digital counts. Normalizing the fitted functions provides us a measured spectral responsivity $R_{\lambda,norm} $ for our Nano-Hyperspec sensor.

We can then use the normalized spectral responsivity curve and the known sensor specifications to obtain irradiance per digital count as

\begin{equation}
    \frac{E}{DC} = \left( \phi_{ref} \cdot \frac{t_{obs}}{t_{ref}} \right) \left(x_{obs}^2 \cdot \theta_{IFOV}^2 \cdot \frac{B_{ref}}{B_{obs}} \cdot R_{\lambda,norm} \right)^{-1},
    \label{eq:MulhollanMethod}
\end{equation}

where the $ref$ and $obs$ subscripts denote acquisition parameters from laboratory (monochrometer) measurement and field measurements respectively. 

Assuming a Lambertian reflector, we calculate sensor reaching spectral radiance as

\begin{equation}
    L_{\lambda} = (DC_i - DC_{dark}) \cdot \frac{E}{DC} \cdot \frac{1}{\pi}.
    \label{eq:SpectralRadiance}
\end{equation}

To convert from spectral radiance to spectral reflectance, we assume the surface is Lambertian and solely illuminated by downwelling irradiance (measured by the ASD) as

\begin{equation}
    \rho_\lambda = L_\lambda \cdot \frac{E_{\lambda,downwell}}{\pi}.
    \label{eq:Reflectance}
\end{equation}

To summarize, we obtain the reflectance spectra as follows:
\begin{itemize}
    \item Calibrate the sensor using monochrometer to obtain the spectral responsivity curve as per Eqn. \ref{eq:responsivity}.
    \item Assume vehicle is Lambertian and sample approximately planar region of interest.
    \item Use Eqn. \ref{eq:MulhollanMethod} to convert ROI data from digital counts to irradiance.
    \item Use Eqn. \ref{eq:SpectralRadiance} and \ref{eq:Reflectance} to compensate the dark current measurements and obtain reflectance spectra.
\end{itemize}




\begin{table}[t]
\centering
\def\arraystretch{1.2}
\begin{tabular}{@{}lll@{}}
\toprule
Variable & Description & Units \\ \midrule
$S_e$ & Sensor Generated Electrons & -\\
$\phi$ & Radiant Flux & \si{\watt} \\
$R_\lambda$ & Spectral Responsivity & \si{\per\watt}\\ 
$M_\lambda$ & Radiant Exitance & \si{\watt\per\meter\squared}\\
$E$ & Irradiance & \si{\watt\per\meter\squared}\\
$\rho_\lambda$ & Spectral Reflectance & \si{\watt\per\meter\squared}\\
$B$ & Spectral Bandwidth & \si{n\meter} \\
$x$ & Ground Sample Distance & \si{\meter} \\
\midrule
$t$ & Time & \si{\second} -\\
$\lambda$ & Wavelength &  \si{n\meter}\\
$h$ & Planck's constant & \si{\joule\second} \\
$c$ & Speed of Light &  \si{\meter\per\second}\\
$\eta$ & Quantum Efficiency & -\\
$\theta$ & Angle of Incidence & \si{rad}\\
$\theta_{IFOV}$ & Instantaneous FOV & \si{sr}\\
\bottomrule
\end{tabular}
\caption{List of all notations used throughout the paper, their description and default units.}
\label{tab:allunits}
\end{table}

\begin{figure}
    \begin{subfigure}{\linewidth}
        \centering
        \includegraphics[width=\linewidth]{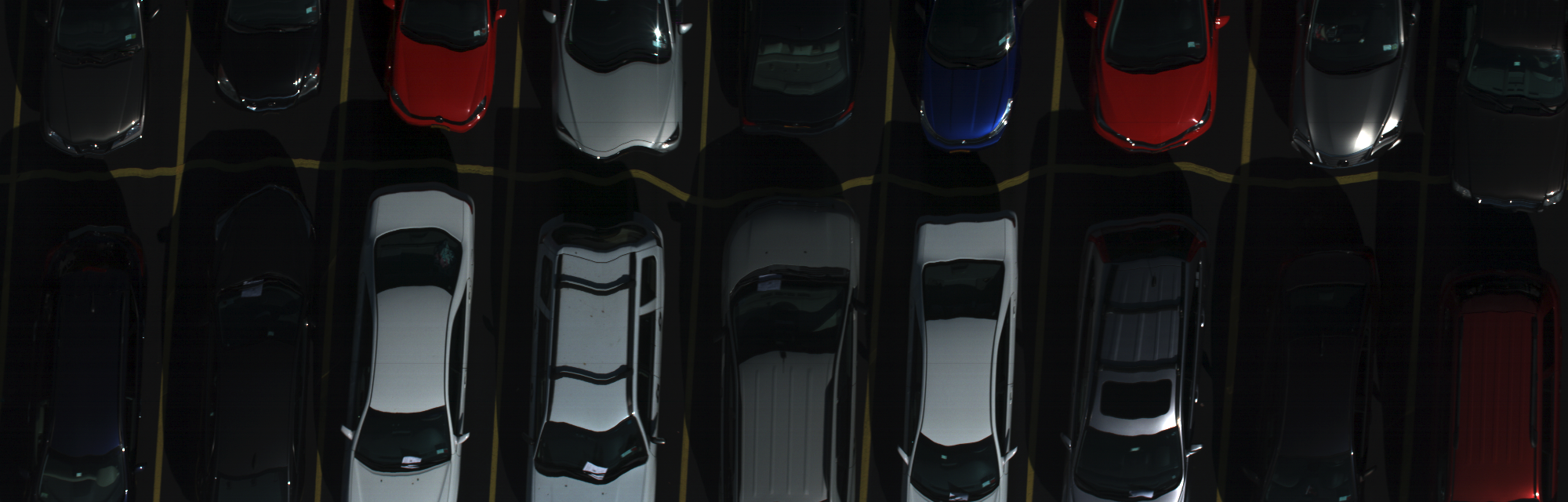}
        \caption{}
        \label{fig:img_raw}
    \end{subfigure}
    \hfill
    \begin{subfigure}{\linewidth}
        \centering
        \includegraphics[width=\linewidth]{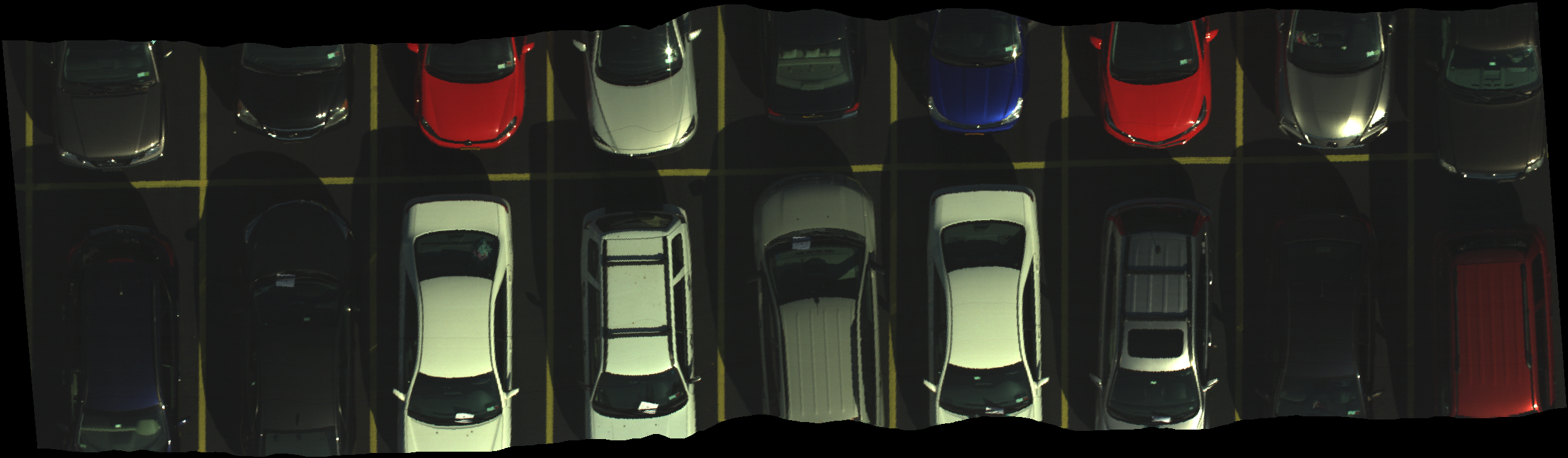}
        \caption{}
        \label{fig:img_or}
    \end{subfigure}
    \caption{We use orthorectification to remove the distortions in \ref{fig:img_raw} and obtain a \ref{fig:img_or}.}
    \label{fig:orthorec}
\end{figure}

\section{Data Processing and Visualization}

\subsection{University Parking Lot}

The drone based imaging platform is susceptible to sharp changes in the collected imagery due to wind-influenced movement (Fig \ref{fig:img_raw}). We use orthorectification based on GPS and IMU data and correct these distortions to obtain a much cleaner image (Fig. \ref{fig:img_or}). Since we operate the sensor in a non-automatic setting (i.e. the gain and integration time does not change with the scene variance), we apply the pipeline described in Sec. \ref{sec:radcalc} after orthorectification. After smoothing out the resultant reflectance signal with a box filter, we import the vehicle spectral curves into DIRSIG as new synthetic paint signatures.

\subsection{ROI Selection}

\label{sec:roiselection}
\begin{figure}
    \centering
    \includegraphics[width = \linewidth]{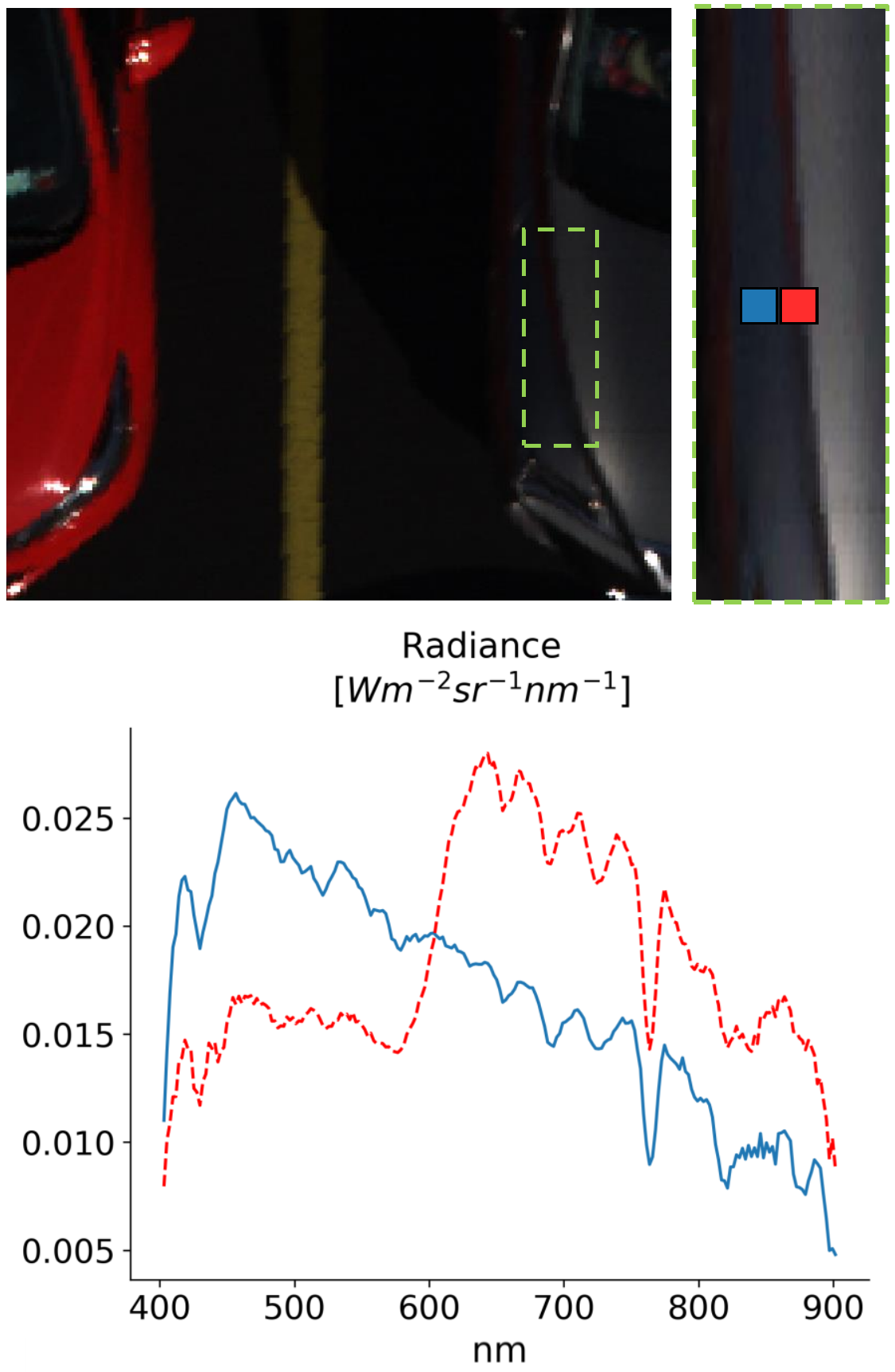}
    \caption{Adjacency effect from the red car onto the gray car. We select two ROIs - the red and blue boxes, and plot the mean spectra with respective colors.}
    \label{fig:adjeffect}
\end{figure}

In our data collect, when radiance due to adjacency effect is proportional to the downwelling radiance for a given pixel, we no longer know how the corresponding pixel is illuminated. Therefore, we cannot obtain accurate vehicle material reflectance(s) in regions that have significant adjacency effects. For example, in Fig. \ref{fig:adjeffect}, a gray vehicle has secondary light reflections from a nearby bright red car. Along contours of the car hood that are pointed towards the red vehicle, we observe a red tinged spectral radiance that is significantly different than the spectral radiance of a nearby region on the gray vehicle hood. 

Vehicle paint is highly specular, and under solar illumination, the vehicle may appear to have bright glinted regions on its surface. Glint is caused by direct reflections from an illumination source. In our parking lot scene, we observed that pixels containing glint are often an order of magnitude brighter than the observed vehicle brightness, which causes our imaging sensor to be saturated. In addition to being brighter, the irradiance spectra from glinted regions on a vehicle do not have the same paint signature as the vehicle paint, and instead resembles the solar spectral irradiance curve. When selecting ROI's on the vehicles surface, it is important to avoid glinted pixels to reduce error in calculating spectral reflectance. An example of glint effects on measured digital counts, radiance, and reflectance is shown in Fig. \ref{fig:Bigfig}. Hence, in order to obtain spectral reflectance curves of objects from our remote imaging platform, we \textit{carefully} select regions on the object that are geometrically planar and are not influenced by secondary reflections from nearby objects.

\subsection{Comparison with SVC Data}
\label{sec:comparisonwithsvc}

\begin{figure}
    \centering
    \includegraphics[width = \linewidth]{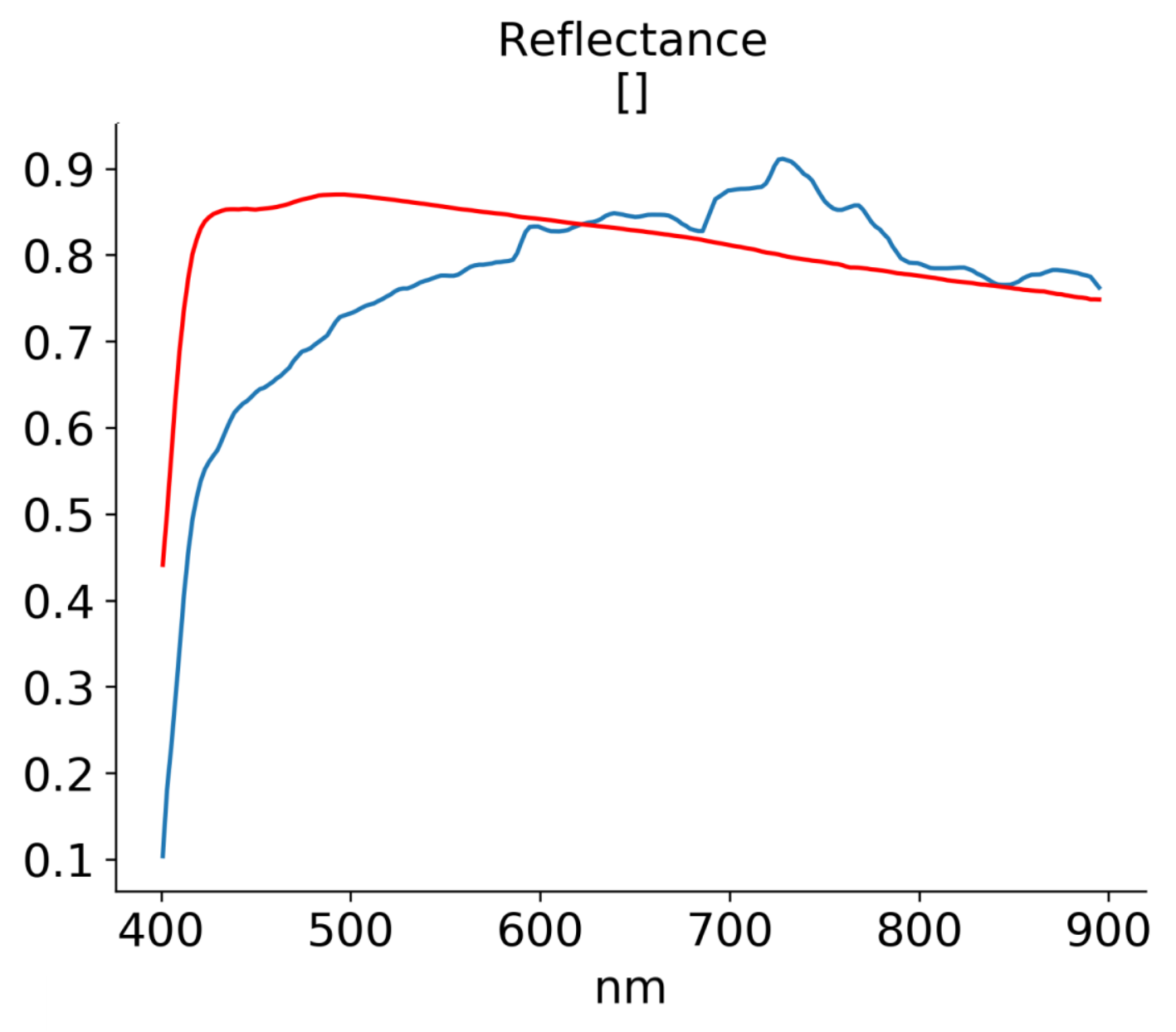}
    \caption{Calculated vehicle paint reflectance from the Nano-HyperSpec (blue) compared with the measured SVC paint reflectance (red). These spectra are both from the white vehicle of interest in Fig. \ref{fig:Bigfig}.}
    \label{fig:reflectance_comparison}
\end{figure}

We then validate our calculated reflectance from the airborne Nano-HyperSpec with reflectance measurements from the SVC spectrometer of known vehicles in the parking lot scene, to determine if our assumptions made in Sec. \ref{sec:Data_Collection} provide reasonable estimates of vehicle paint reflectance curves. Following equations laid out in Sec. \ref{sec:radcalc}, we show the calculated reflectance plotted alongside the ground truth reflectance (SVC spectrometer) in Fig. \ref{fig:reflectance_comparison}.

Considering the radiometric simplifications utilized to obtain vehicle paint reflectance spectra, the results reasonably agree with the ground truth spectra. The discrepancies between the two curves could be due to multiple reasons, including 1) adjacency effect, 2) the full sky hemisphere of downwelling irradiance is not visible to the selected ROI due to occlusion, and 3) the vehicle paint BRDF is not Lambertian and is not perfectly planar within the ROI.

\subsection{Hyperspectral Scene Rendering}

\begin{figure}
    \centering
    \includegraphics[width = \linewidth]{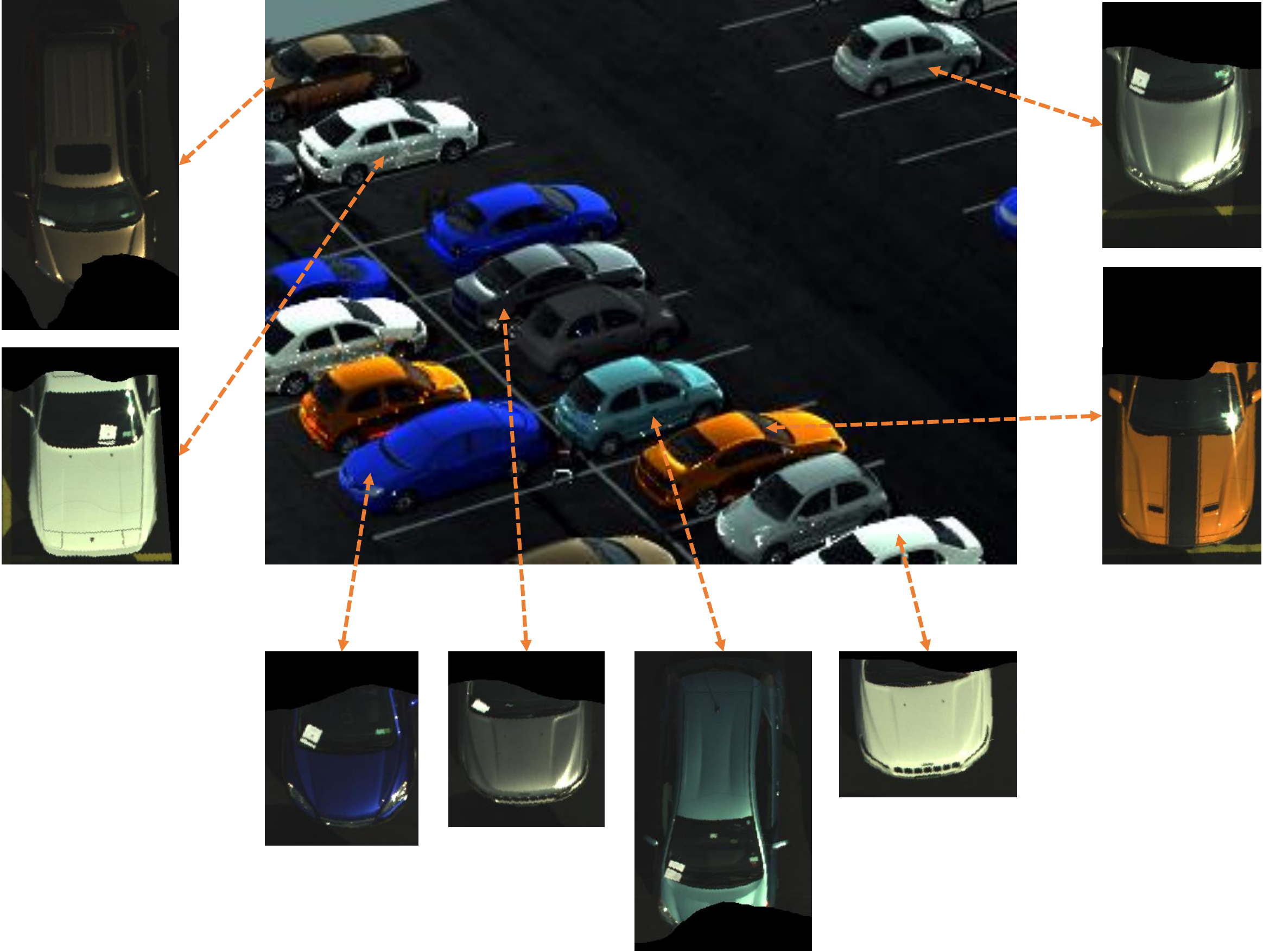}
    \caption{Center - Synthetic parking lot scene generated by DIRSIG with vehicles exhibiting paints specifically obtained via our approach. Surround - We also show some of the vehicles used for sampling the paint instances.}
    \label{fig:DIRSIGpaints}
\end{figure}

After calibrating the car paint spectra and porting it into DIRSIG, we use a predefined parking lot scene \footnote{\href{http://www.dirsig.org/docs/demos/index.html}{http://www.dirsig.org/docs/demos/index.html}} in DIRSIG and change the material properties to use our set of paint signatures for the vehicles instead of the predefined set. For preliminary testing, we select 14 vehicle reflectance spectra for simulations. Fig. \ref{fig:DIRSIGpaints} shows a snapshot of the rendered scene along with new set of paint signatures and their corresponding real image counterparts that were used for extraction. The image rendered is significantly low in resolution as compared to the real image from which colors are sampled due to the altitude difference of the sensor. 

\begin{figure}
    \centering
    \begin{subfigure}[t]{.23\textwidth}
        \centering
        \includegraphics[width=\linewidth]{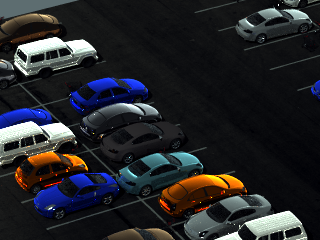}
        \caption{}
        \label{fig:scene_simple}
    \end{subfigure}
    \hfill
    \begin{subfigure}[t]{.23\textwidth}
        \centering
        \includegraphics[width=\linewidth]{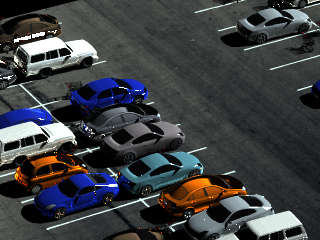}
        \caption{}
        \label{fig:scene_mlsatm}
    \end{subfigure}
    
    \medskip
    
    \begin{subfigure}[t]{.48\textwidth}
        \centering
        \includegraphics[width=\linewidth]{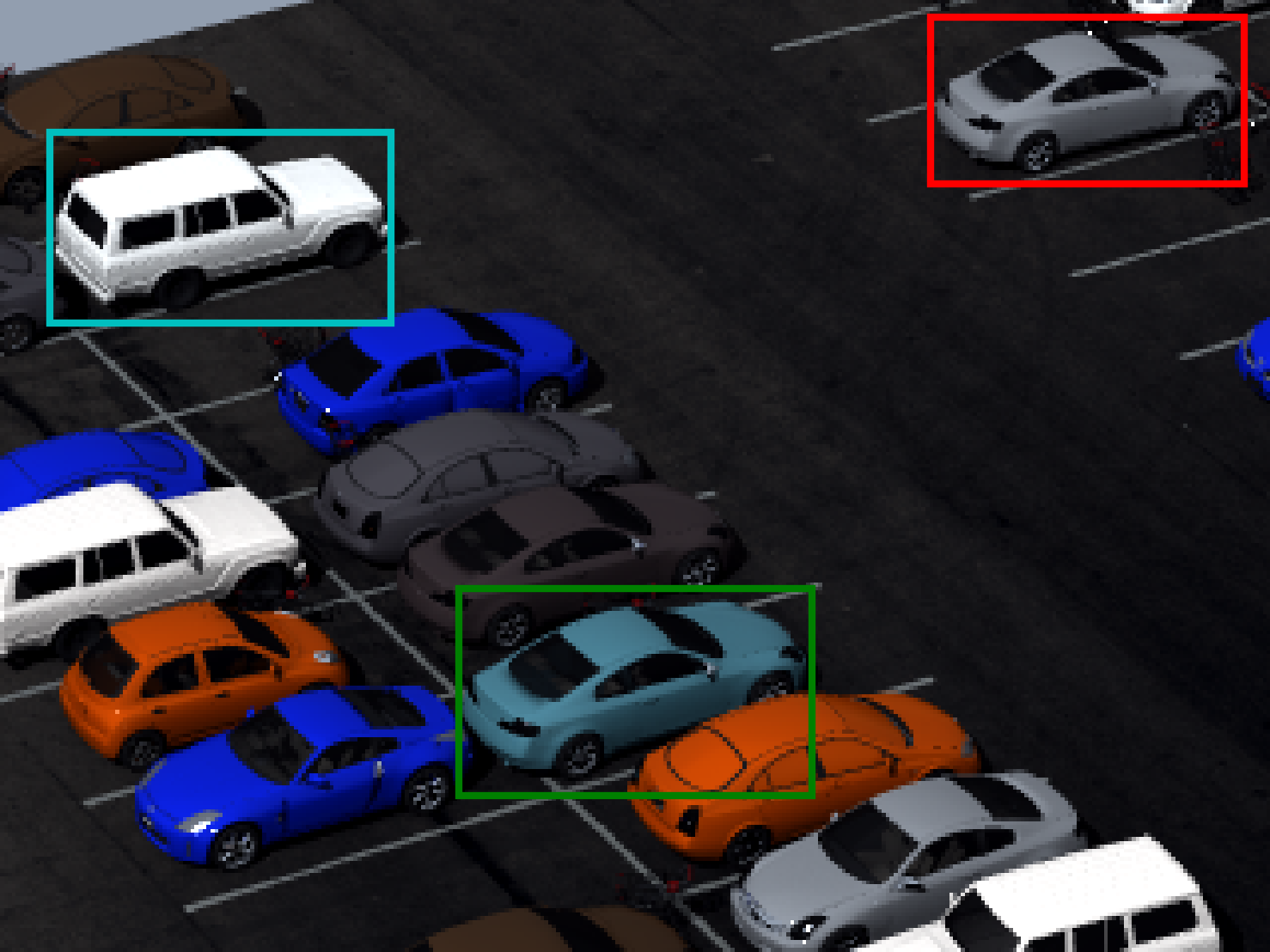}
        \caption{}
        \label{fig:scene_cloudy}
    \end{subfigure}
    \medskip
    \begin{subfigure}[t]{.42\textwidth}
        \centering
        \includegraphics[width=\linewidth]{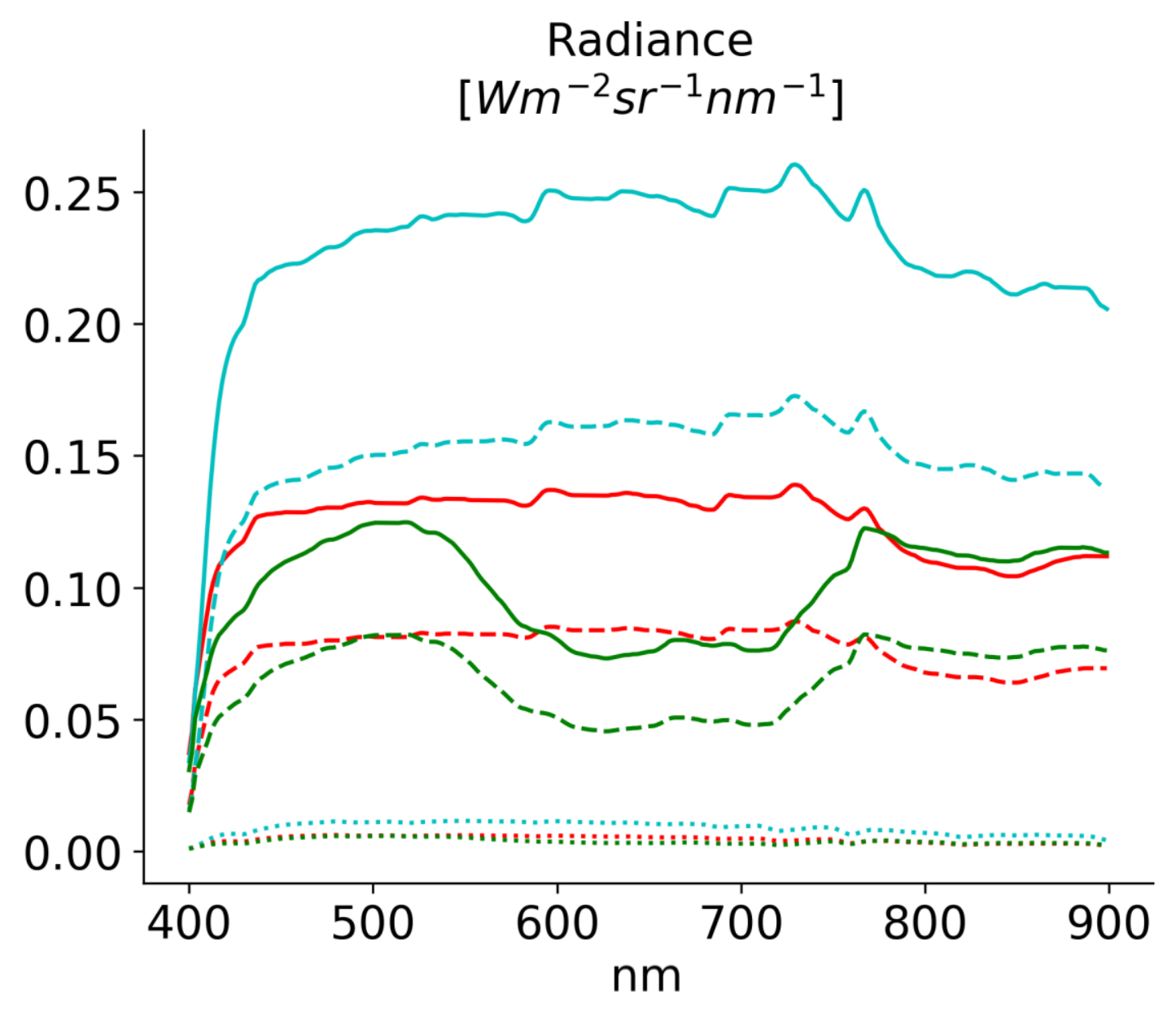}
        \caption{}
        \label{fig:scene_allgraphs}
    \end{subfigure}

    \caption{Same parking lot scene captured under three atmospheric conditions - at noon (\ref{fig:scene_simple}), near sunset (\ref{fig:scene_mlsatm}) and under clouds (\ref{fig:scene_cloudy}). We plot the mean spectrum of pixels sampled from selected cars (marked in \ref{fig:scene_cloudy}) in \ref{fig:scene_allgraphs} with different line styles: noon - solid, near sunset - dotted, cloudy - dashed.}
    \label{fig:dirsigillumns}
    
\end{figure}

Another benefit of creating simulated hyperspectral scenes is that we have full knowledge of the ground truth properties of objects placed in the scene. We then can observe how the spectral radiance reaching our sensor is influenced by angle, illumination spectra, atmosphere, as well as platform and object motion. Through many simulations of different observation conditions, we can measure  in observed spectral radiance that corresponds to an object of known reflectance properties, to aid in future hyperspectral tasks - involving object detection and re-identification. Fig. \ref{fig:dirsigillumns} shows another instance of the parking lot scene with varied vehicle paints under three atmospheric conditions - at noon, near sunset and under clouds \footnote{we add the individual plots in the supplemental}. We observe stark differences at all three timestamps - signatures under cloud tend to have less intensity compared to the ones at noon depending on the base paint color, and signatures during evening are all at very less intensity irrespective of the paint.

\section{Discussion and Future Work}

We have shown a radiometrically calibrated airborne hyperspectral platform operating simultaneously with a spectrometer measuring downwelling irradiance can calculate reflectance spectra of vehicle paints from manually selected ROIs on the vehicle surface. We select pixels from approximately planar areas on the vehicle body where we assume the dominant source of illumination is the downwelling irradiance, thus avoiding pixels that are significantly corrupted due to adjacency effect or glint or shadow. 

These reflectance spectra are utilized in DIRSIG to generate simulated hyperspectral imagery, with the potential to create dense simulated urban scenes where each vehicle has a unique spectral signature. These simulations allow for the construction of multiple spectral radiance curves that correspond to the same object material (Fig. \ref{fig:dirsigillumns}), generating examples of how an object can look different when there is local or radiometric changes in the scene. We believe that our collection of paint spectra can be used with DIRSIG for future work in multiple areas of remote sensing, including but not limited to data augmentation for hyperspectral object detection from multiple platforms.

\ifthenelse{\boolean{ack}}
{\section*{Acknowledgements}
This work has been supported by the Dynamic Data Driven Applications Systems Program, Air Force Office of Scientific Research under Grant FA9550-19-1-0021. We thank Carl Salvaggio, Nanette Salvaggio, Emmett Ientilucci, Nina Raqueno and Tania Kleynhans for feedback throughout the drafting phase of this paper.}

{\small
\bibliographystyle{ieee_fullname}
\bibliography{egbib}

\begin{thebibliography}{10}\itemsep=-1pt

\bibitem{bromley1994signature}
Jane Bromley, Isabelle Guyon, Yann LeCun, Eduard S{\"a}ckinger, and Roopak
  Shah.
\newblock Signature verification using a" siamese" time delay neural network.
\newblock In {\em Advances in neural information processing systems}, pages
  737--744, 1994.

\bibitem{imagenet_cvpr09}
J. Deng, W. Dong, R. Socher, L.-J. Li, K. Li, and L. Fei-Fei.
\newblock {ImageNet: A Large-Scale Hierarchical Image Database}.
\newblock In {\em CVPR09}, 2009.

\bibitem{dosovitskiy2017carla}
Alexey Dosovitskiy, German Ros, Felipe Codevilla, Antonio Lopez, and Vladlen
  Koltun.
\newblock Carla: An open urban driving simulator.
\newblock {\em arXiv preprint arXiv:1711.03938}, 2017.

\bibitem{goodenough2017dirsig5}
Adam~A Goodenough and Scott~D Brown.
\newblock Dirsig5: next-generation remote sensing data and image simulation
  framework.
\newblock {\em IEEE Journal of Selected Topics in Applied Earth Observations
  and Remote Sensing}, 10(11):4818--4833, 2017.

\bibitem{han2017efficient}
Sanghui Han, Alex Fafard, John Kerekes, Michael Gartley, Emmett Ientilucci,
  Andreas Savakis, Charles Law, Jason Parhan, Matt Turek, Keith Fieldhouse,
  et~al.
\newblock Efficient generation of image chips for training deep learning
  algorithms.
\newblock In {\em Automatic Target Recognition XXVII}, volume 10202, page
  1020203. International Society for Optics and Photonics, 2017.

\bibitem{he2017mask}
Kaiming He, Georgia Gkioxari, Piotr Doll{\'a}r, and Ross Girshick.
\newblock Mask r-cnn.
\newblock In {\em Proceedings of the IEEE international conference on computer
  vision}, pages 2961--2969, 2017.

\bibitem{he2016deep}
Kaiming He, Xiangyu Zhang, Shaoqing Ren, and Jian Sun.
\newblock Deep residual learning for image recognition.
\newblock In {\em Proceedings of the IEEE conference on computer vision and
  pattern recognition}, pages 770--778, 2016.

\bibitem{ientilucci2003advances}
Emmett~J Ientilucci and Scott~D Brown.
\newblock Advances in wide-area hyperspectral image simulation.
\newblock In {\em Targets and Backgrounds IX: Characterization and
  Representation}, volume 5075, pages 110--121. International Society for
  Optics and Photonics, 2003.

\bibitem{kaputa2019mx}
Daniel~S Kaputa, Timothy Bauch, Carson Roberts, Don McKeown, Mark Foote, and
  Carl Salvaggio.
\newblock Mx-1: A new multi-modal remote sensing uas payload with high accuracy
  gps and imu.
\newblock In {\em 2019 IEEE Systems and Technologies for Remote Sensing
  Applications Through Unmanned Aerial Systems (STRATUS)}, pages 1--4. IEEE,
  2019.

\bibitem{kemker2018algorithms}
Ronald Kemker, Carl Salvaggio, and Christopher Kanan.
\newblock Algorithms for semantic segmentation of multispectral remote sensing
  imagery using deep learning.
\newblock {\em ISPRS Journal of Photogrammetry and Remote Sensing}, 145:60--77,
  2018.

\bibitem{kolb2016digital}
Kimberly~E Kolb, S~Choi Hee-sue, Balvinder Kaur, Jeffrey~T Olson, Clayton~F
  Hill, and James~A Hutchinson.
\newblock Digital imaging and remote sensing image generator (dirsig) as
  applied to nvesd sensor performance modeling.
\newblock In {\em Infrared Imaging Systems: Design, Analysis, Modeling, and
  Testing XXVII}, volume 9820, page 982019. International Society for Optics
  and Photonics, 2016.

\bibitem{lin2014microsoft}
Tsung-Yi Lin, Michael Maire, Serge Belongie, James Hays, Pietro Perona, Deva
  Ramanan, Piotr Doll{\'a}r, and C~Lawrence Zitnick.
\newblock Microsoft coco: Common objects in context.
\newblock In {\em European conference on computer vision}, pages 740--755.
  Springer, 2014.

\bibitem{meyers2002incorporation}
Jason~P Meyers, John~R Schott, and Scott~D Brown.
\newblock Incorporation of polarization into the dirsig synthetic image
  generation model.
\newblock In {\em Imaging Spectrometry VIII}, volume 4816, pages 132--143.
  International Society for Optics and Photonics, 2002.

\bibitem{moorhead2001cameo}
Ian~R Moorhead, Marilyn~A Gilmore, Alexander~W Houlbrook, David~E Oxford,
  David~R Filbee, Colin~A Stroud, George Hutchings, and Albert Kirk.
\newblock Cameo-sim: a physics-based broadband scene simulation tool for
  assessment of camouflage, concealment, and deception methodologies.
\newblock {\em Optical Engineering}, 40, 2001.

\bibitem{rahman2018siamese}
Faiz Rahman, Bhavan Vasu, Jared Van~Cor, John Kerekes, and Andreas Savakis.
\newblock Siamese network with multi-level features for patch-based change
  detection in satellite imagery.
\newblock In {\em 2018 IEEE Global Conference on Signal and Information
  Processing (GlobalSIP)}, pages 958--962. IEEE, 2018.

\bibitem{rangnekar2019aerorit}
Aneesh Rangnekar, Nilay Mokashi, Emmett Ientilucci, Christopher Kanan, and
  Matthew~J Hoffman.
\newblock Aerorit: A new scene for hyperspectral image analysis.
\newblock {\em arXiv preprint arXiv:1912.08178}, 2019.

\bibitem{Richtsmeier2001A3R}
Steven~C. Richtsmeier, Alexander Berk, Lawrence~S. Bernstein, and Steven~M.
  Adler-Golden.
\newblock A 3-dimensional radiative-transfer hyperspectral image simulator for
  algorithm validation.
\newblock 2001.

\bibitem{savva2019habitat}
Manolis Savva, Abhishek Kadian, Oleksandr Maksymets, Yili Zhao, Erik Wijmans,
  Bhavana Jain, Julian Straub, Jia Liu, Vladlen Koltun, Jitendra Malik, et~al.
\newblock Habitat: A platform for embodied ai research.
\newblock In {\em Proceedings of the IEEE International Conference on Computer
  Vision}, pages 9339--9347, 2019.

\bibitem{dirsig1}
J.R. Schott, S.D. Brown, R.V. Raqueño, H.N. Gross, and G. Robinson.
\newblock An advanced synthetic image generation model and its application to
  multi/hyperspectral algorithm development.
\newblock {\em Canadian Journal of Remote Sensing}, 25(2):99--111, 1999.

\bibitem{simonyan2014very}
Karen Simonyan and Andrew Zisserman.
\newblock Very deep convolutional networks for large-scale image recognition.
\newblock {\em arXiv preprint arXiv:1409.1556}, 2014.

\bibitem{uzkent2016integrating}
Burak Uzkent, Matthew~J Hoffman, and Anthony Vodacek.
\newblock Integrating hyperspectral likelihoods in a multidimensional
  assignment algorithm for aerial vehicle tracking.
\newblock {\em IEEE Journal of Selected Topics in Applied Earth Observations
  and Remote Sensing}, 9(9):4325--4333, 2016.

\bibitem{uzkent2018tracking}
Burak Uzkent, Aneesh Rangnekar, and Matthew~J Hoffman.
\newblock Tracking in aerial hyperspectral videos using deep kernelized
  correlation filters.
\newblock {\em IEEE Transactions on Geoscience and Remote Sensing},
  57(1):449--461, 2018.

\bibitem{wrenninge2018synscapes}
Magnus Wrenninge and Jonas Unger.
\newblock Synscapes: A photorealistic synthetic dataset for street scene
  parsing.
\newblock {\em arXiv preprint arXiv:1810.08705}, 2018.

\bibitem{wymann2015torcs}
Bernhard Wymann, Christos Dimitrakakis, Andrew Sumner, Eric Espi{\'e}, and
  Christophe Guionneau.
\newblock Torcs, the open racing car simulator.
\newblock 2015.

\bibitem{zahidi2020end}
Usman~A Zahidi, Peter~WT Yuen, Jonathan Piper, and Peter~S Godfree.
\newblock An end-to-end hyperspectral scene simulator with alternate adjacency
  effect models and its comparison with cameosim.
\newblock {\em Remote Sensing}, 12(1):74, 2020.

\end{thebibliography}
}

\ifthenelse{\boolean{combined}}{
\clearpage
\begin{center}
    {\Large Supplemental Material \normalsize}
\end{center}
\beginsupplement

We plot the mean spectra sampled from the roof of the vehicles shown in Fig. \ref{fig:objbbox} under three atmospheric conditions - at noon (Fig. \ref{fig:detail-simple}, near sunset \ref{fig:detail-mlsatm} and under clouds \ref{fig:detail-cloudy}). We modeled a simple atmosphere with uniform spectral absorption for Fig. \ref{fig:detail-simple} and Fig. \ref{fig:detail-cloudy} and a realistic atmosphere with water and oxygen absorption bands for Fig. \ref{fig:detail-mlsatm}. The most prominent absorption band in the VNIR spectral range is located at $\approx$ 760nm, causing a visible drop in  the vehicle spectral radiance curves in Fig. \ref{fig:detail-mlsatm}.

\begin{figure}[t]
        \centering
        \includegraphics[width=\linewidth]{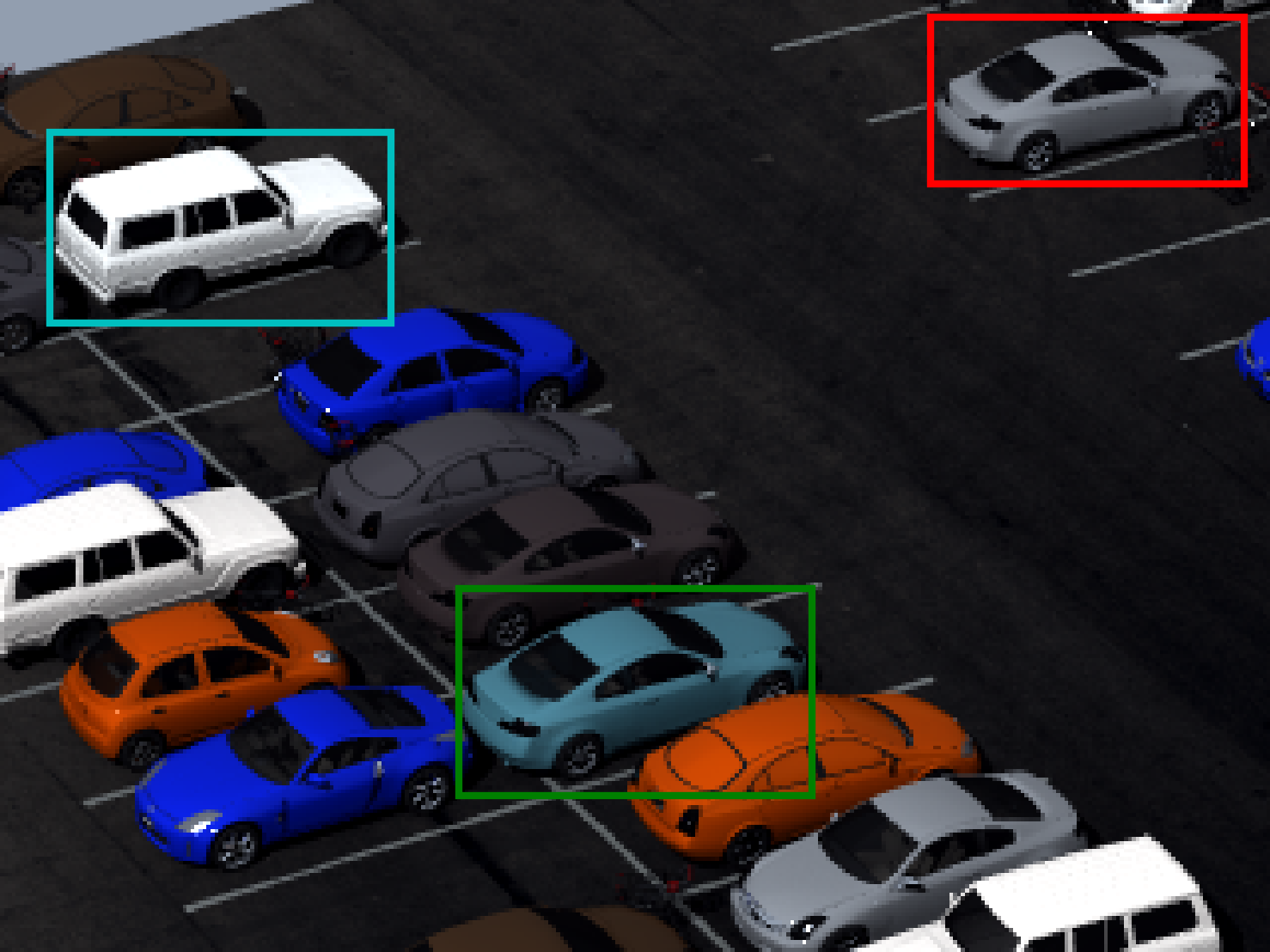}
        \caption{Bounding boxes on vehicles on interest for Figs. \ref{fig:detail-simple}, \ref{fig:detail-mlsatm} and \ref{fig:detail-cloudy}.}
        \label{fig:objbbox}
\end{figure}

\begin{figure}[t]
        \centering
        \includegraphics[width=\linewidth]{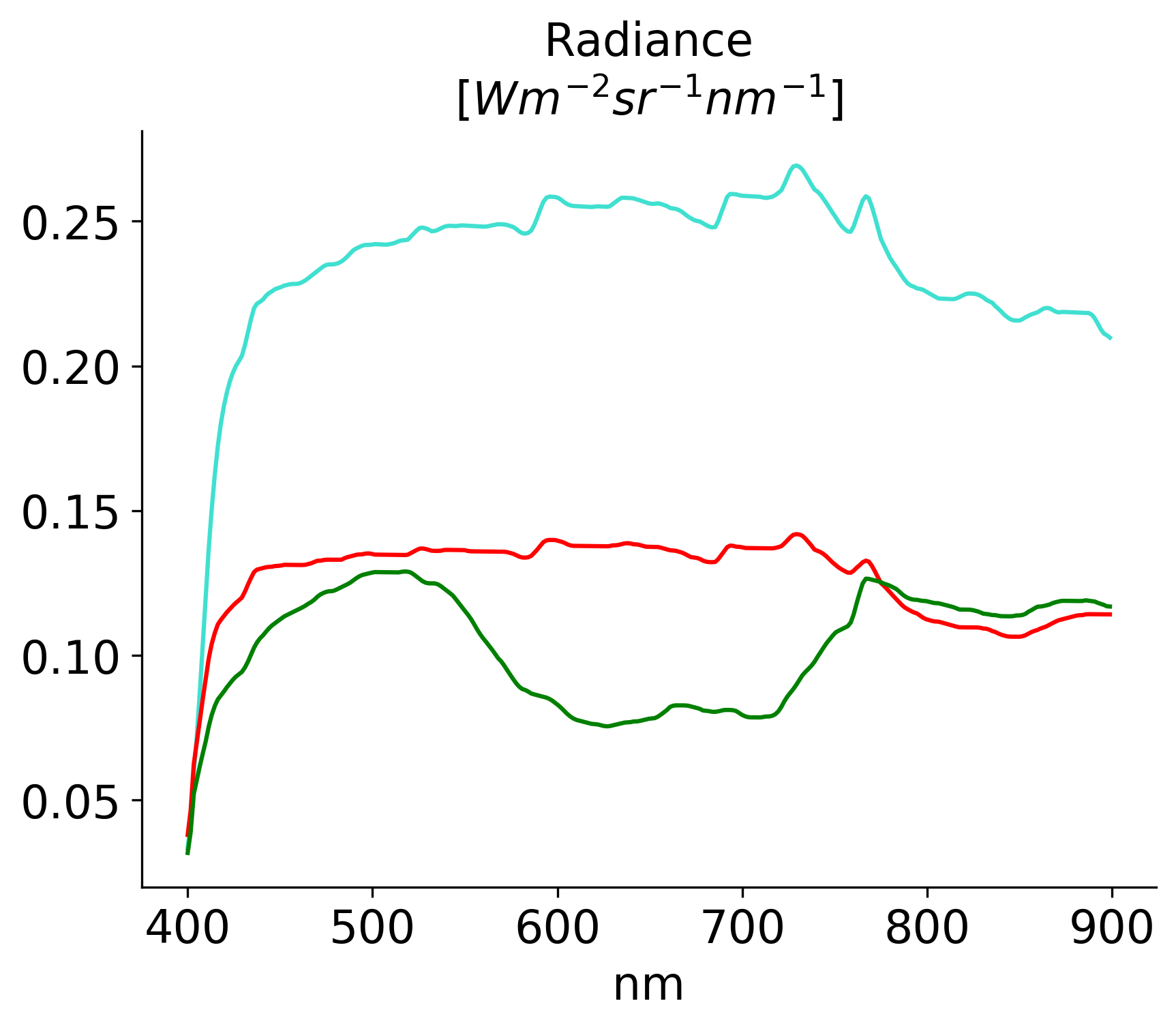}
        \caption{Mean spectra of vehicles shown in Fig. \ref{fig:objbbox} at noon.}
        \label{fig:detail-simple}
\end{figure}

\begin{figure}[t]
        \centering
        \includegraphics[width=\linewidth]{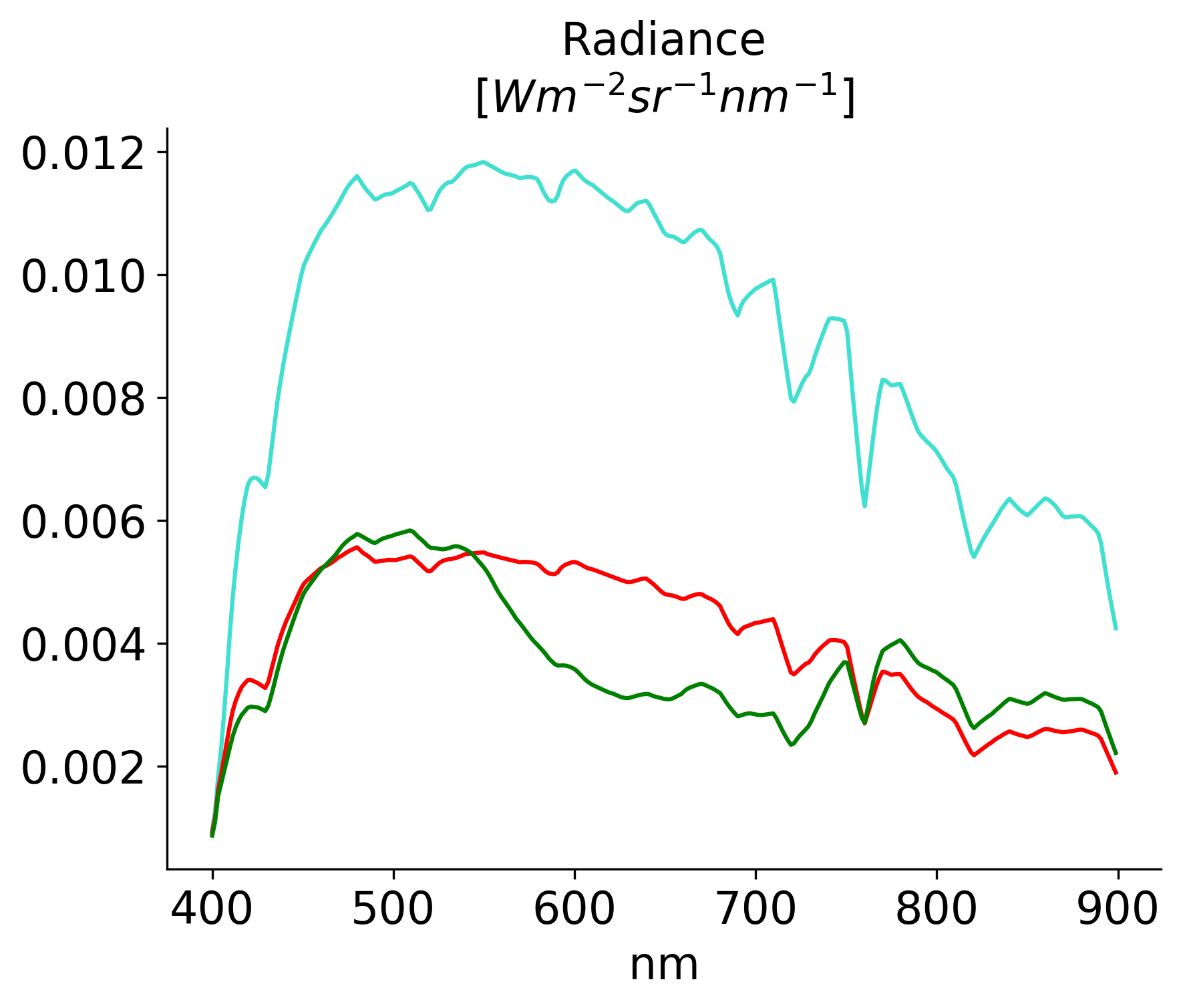}
        \caption{Mean spectra of vehicles shown in Fig. \ref{fig:objbbox} near sunset.}
        \label{fig:detail-mlsatm}
\end{figure}

\begin{figure}[t]
        \centering
        \includegraphics[width=\linewidth]{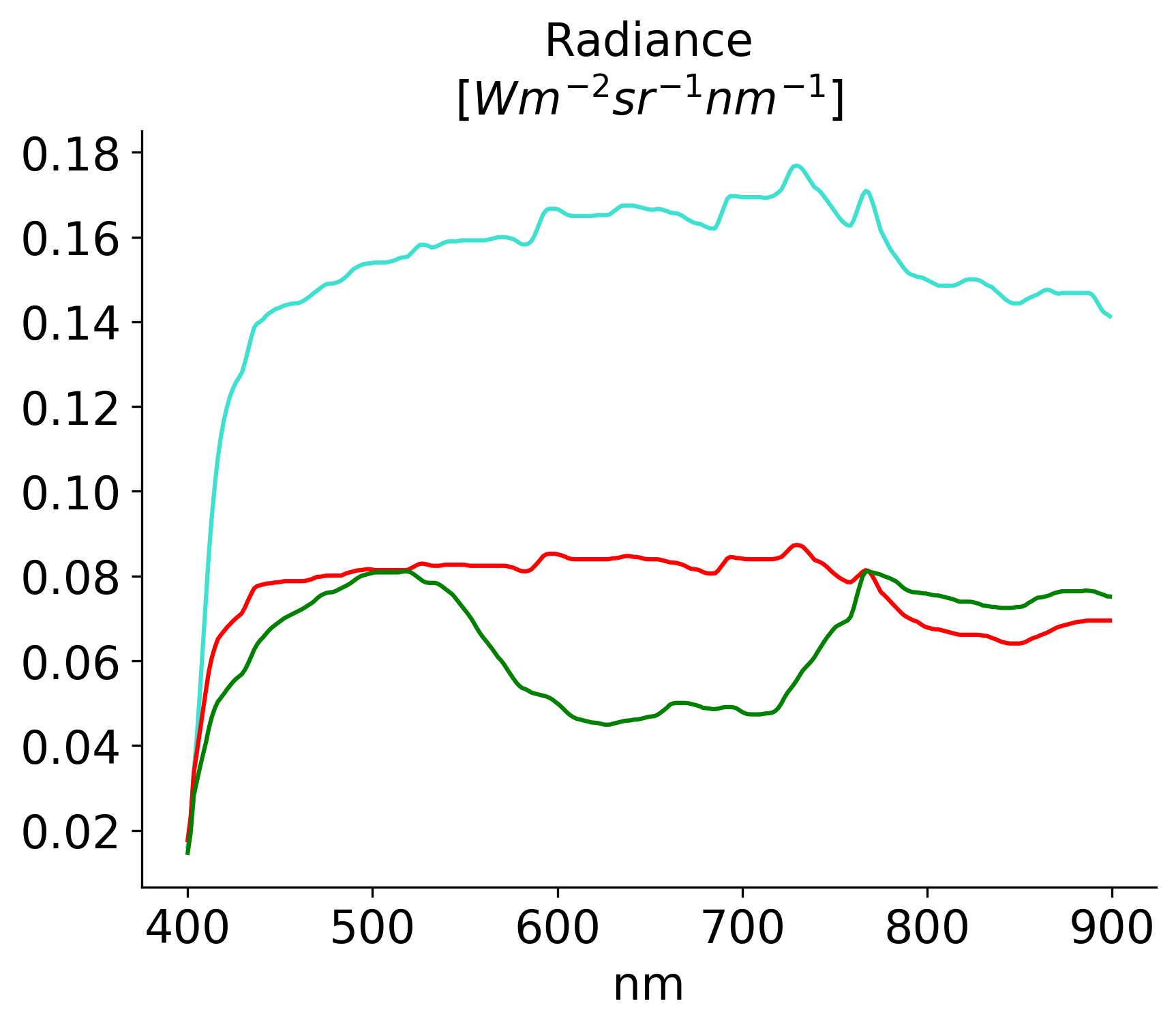}
        \caption{Mean spectra of vehicles shown in Fig. \ref{fig:objbbox} under clouds.}
        \label{fig:detail-cloudy}
\end{figure}

}

\end{document}